# Projective spin adaptation for the exact diagonalization of isotropic spin clusters

Shadan Ghassemi Tabrizi, Thomas D. Kühne

*Center for Advanced Systems Understanding (CASUS),*
*Am Untermarkt 20, 02826 Görlitz, Germany*
s.ghassemi-tabrizi@hzdr.de

**Abstract.** Spin Hamiltonians, like the Heisenberg model, are used to describe magnetic properties of exchange-coupled molecules and solids. For finite clusters, physical quantities such as heat capacities, magnetic susceptibilities or neutron-scattering spectra, can be calculated based on energies and eigenstates obtained by exact diagonalization (ED). Utilizing spin-rotational symmetry SU(2) to factor the Hamiltonian with respect to total spin $S$ facilitates ED, but the conventional approach to spin-adapting the basis is more intricate than selecting states with a given magnetic quantum number $M$ (the spin $z$-component), as it relies on irreducible tensor-operator techniques and spin-coupling coefficients. Here, we present a simpler technique based on applying a spin projector to uncoupled basis states. As an alternative to Löwdin's projection operator, we consider a group-theoretical formulation of the projector, which can be evaluated either exactly or approximately using an integration grid. An important aspect is the choice of uncoupled basis states. We present an extension of Löwdin's theorem for $s = \frac{1}{2}$ to arbitrary local spin quantum numbers $s$, which allows for the direct selection of configurations that span a complete, linearly independent basis in an $S$ sector upon spin projection. We illustrate the procedure with a few examples.

## 1. Introduction

With the discovery of molecular magnetism in the early 1990s [1,2] and the subsequent synthesis of a variety of exchange-coupled clusters, theoretical modeling based on multi-spin Hamiltonians has come to the forefront of analyzing the intriguing properties of multinuclear transition-metal complexes [3–5]. The leading term is usually of Heisenberg type, $\hat{H} = \sum_{i<j} J_{ij}\hat{\mathbf{s}}_i \cdot \hat{\mathbf{s}}_j$ [3], and this model allows to describe thermodynamic properties like magnetic susceptibilities or heat capacities over a wide temperature range; anisotropic contributions become significant only at low temperatures for systems with a magnetic ground state (total spin $S > 0$). Besides, for an individual spin multiplet (comprising states with $\hat{S}_z$ eigenvalues $M = -S, -S+1, ..., +S$) anisotropic terms that lift the $(2S+1)$-fold degeneracy (zero-field splitting) can often be treated perturbatively [3,6,7]. Thus, techniques for the efficient solution of isotropic spin Hamiltonians, which may also include other interaction terms, like biquadratic exchange, are of particular importance.

In its most basic form, exact diagonalization (ED) quickly encounters limits due to the rapid growth of the Hilbert-space dimension, $D = \prod_{i=1}^{N}(2s_i + 1)$, with the number $N$ of sites, where $s_i$ denotes a local spin quantum number. Therefore, it becomes necessary to factorize the Hamiltonian according to different irreducible representations of the symmetry group [8]. This not only reduces computation times but also decreases memory requirements, which are usually the limiting factor in practice. The symmetries that can be utilized are the spin-rotational symmetry, which assigns a quantum number $S$ to each level, and the point-group symmetry (PG), which manifests as symmetry under permutations of sites [8]. Here, we will focus on spin symmetry.

It is straightforward to use only the $z$-component $\hat{S}_z$ of the total spin by working in a basis of uncoupled states $|m_1, ..., m_N\rangle$ defined by a set of $\hat{s}_{i,z}$ eigenvalues $m_i$ of the individual sites, with a selected value $M = \sum_i m_i$. On the other hand, adapting the basis to have definite spin $S$ leads to smaller matrices, but the conventional procedure is more complex, as it relies on irreducible tensor-operator (ITO) techniques, where successively coupled states are decoupled in the calculation of Hamiltonian matrix elements using Wigner-9$j$ symbols. This scheme is explained in textbooks [3,9] and it is implemented in the `MAGPACK` [10] and PHI [11] programs. For a detailed recent account with numerous examples, see Ref. [12].

Our present strategy of spin-projecting uncoupled states $|m_1, ..., m_N\rangle$ is simpler, as it does not require spin-coupling or ITO techniques. To the best of our knowledge, such an approach has in only two instances been applied to Heisenberg clusters: Bernu et al. [13] used Löwdin's projector (see Theory section) in conjunction with spatial-symmetry adaptation for Lanczos ED of triangular-lattice sections containing up to 36 spin-1/2 sites, and by a similar procedure we recently derived eigenvalues in symbolic form for particularly small systems [14]. Here, we are concerned with spin symmetry only, where two aspects come to the forefront: the selection of configurations and the practical realization of projection. In the following Theory section, we initially address the former question, which for $s = \frac{1}{2}$ has already been answered by Löwdin's theorem [15–18]. We extend this theorem to arbitrary $s$ and then discuss additional aspects of spin projection, emphasizing that a group-theoretical projector can be evaluated directly [17], without the need to discretize the relevant integral. This offers numerical advantages over Löwdin's projector. In the subsequent Results and Discussion section, we illustrate the selection of uncoupled basis functions according to our extension of Löwdin's theorem and investigate the numerical accuracy of different projection methods through examples.

## 2. Theory

The essence of the present scheme for the spin factorization of the Hamiltonian consists in applying a projector $\hat{P}_S$ to configurations $|m_1,...,m_N\rangle$ with a definite $z$-projection $M = \sum_i m_i$. As mentioned in the Introduction, Löwdin's operator [19], Eq. (1), has in a few cases [13,14] been used in the diagonalization of Heisenberg models:

$$\hat{P}_S = \prod_{l \neq S} \frac{\hat{\mathbf{S}}^2 - l(l+1)}{S(S+1) - l(l+1)} \quad . \tag{1}$$

Applying $\hat{P}_S$ to a state with definite $M$ affords a pure-spin state $|S,M\rangle$, or the state is annihilated if it has no contributions of spin $S$.

Before discussing practical aspects and presenting an alternative formulation of $\hat{P}_S$, we want to address the question of how to select functions $|m_1,...,m_N\rangle$ such that, upon application of $\hat{P}_S$, a complete and linearly independent set spanning the respective $S$ sector is generated. The total number of these multiplets is denoted $\dim(S)$. If one chooses $\dim(S)$ configurations randomly, the projected functions will generally exhibit linear dependencies. This problem has long been solved for $s = \frac{1}{2}$ [15]. Proofs of the underlying so-called Löwdin theorem [18] were presented by Gershgorn [16] and Pauncz [17]. This theorem states that, for the application of $\hat{P}_S$ in the $M = S$ sector of an $s = \frac{1}{2}$ system, only those uncoupled states should be selected where the cumulative sum of the $m_i$ is never negative, i.e., $\sum_{i}^{n} m_i \geq 0$ for $n = 1, ..., N$. For a system of twelve spin-1/2 centers, we illustrate this in Figure 1 for $S = 0, 1, 2$.

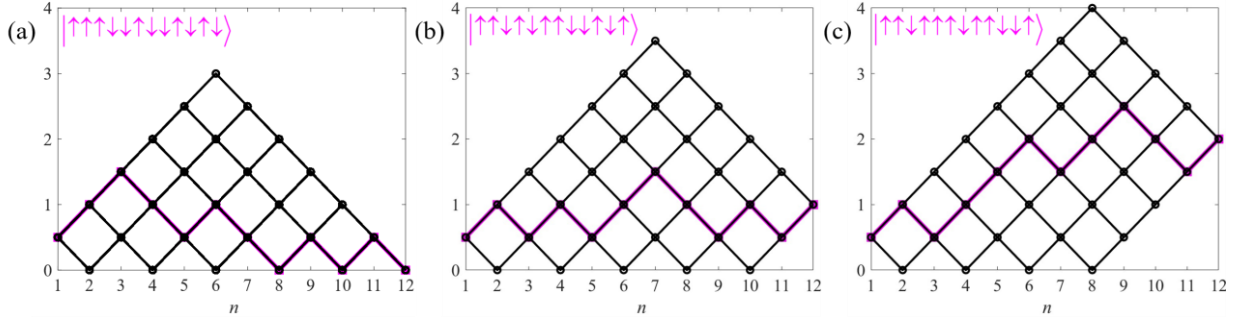

Figure 1: Cumulative sum of the local $z$-projections, $\sum_{i=1}^{n} m_i$, as a function of $n$, for twelve spin-1/2 centers. Each path in the diagram, moving from left to right, represents a configuration selected by Löwdin's theorem, i.e., $\sum_{i=1}^{n} m_i \geq 0$ for all $n$. With respect to $\hat{P}_S$ ($S = M$), these configurations form a complete, linearly independent (though nonorthogonal) set. In each panel (a: $S = 0$, b: $S = 1$, c: $S = 2$), an arbitrary configuration is highlighted and defined for illustration.

To our knowledge, the selection of configurations in $s > \frac{1}{2}$ systems to ensure the linear independence of spin-projected states has not been addressed before. In the following, we explain how Löwdin's theorem can be intuitively extended from $s = \frac{1}{2}$ to any $s$, including systems with different local spin on different sites. The idea leading to this generalization is to embed $|s_i, m_i\rangle$ in a larger space where the site is decomposed into $2s_i$ auxiliary spin-1/2 degrees of freedom $\hat{\kappa}_{i,a}$, see Eq. (2).

$$\hat{\mathbf{s}}_i = \sum_{a=1}^{2s_i} \hat{\kappa}_{i,a} \qquad (2)$$

A state with maximum total spin is symmetric under all permutations of its constituents. For instance, $|s_i = 2, m_i = 1\rangle$ can be represented as a symmetric linear combination of six spin-1/2 configurations related by all possible permutations, see Eq. (3).

$$|s_i = 2, m_i = 1\rangle = \frac{1}{\sqrt{6}}\left(|\uparrow\uparrow\downarrow\downarrow\rangle + |\uparrow\downarrow\uparrow\downarrow\rangle + |\uparrow\downarrow\downarrow\uparrow\rangle + |\downarrow\uparrow\uparrow\downarrow\rangle + |\downarrow\uparrow\downarrow\uparrow\rangle + |\downarrow\downarrow\uparrow\uparrow\rangle\right) \qquad (3)$$

The number of linearly independent states for a given $S$ in the spin-1/2 representation is greater than (for $S < S_{\max}$) or equal to [for $S = S_{\max} = \sum_i s_i$, $\dim(S_{\max}) = 1$] the corresponding number in the actual system (in the latter, $\hat{\kappa}_{i,a}$ are coupled to yield $s_i$). Thus, by first selecting

spin-1/2 configurations according to Löwdin's theorem, then applying $\hat{P}_S$ and finally eliminating contributions with local spin quantum numbers below their respective maximum value $s_i$, an overcomplete basis would be spanned, which we like to avoid.

For the sake of the argument, a local projector is defined in Eq. (4), although this need not be constructed in practical calculations:

$$\hat{p}_{s_i} = \prod_{l \neq s_i} \frac{\left(\sum_a \hat{\kappa}_{i,a}\right)^2 - l(l+1)}{s_i(s_i+1) - l(l+1)} \quad . \tag{4}$$

A state in the spin-1/2 space is mapped onto the smaller space with local spins $\{s_i\}$ through application of the product $\hat{p}$ of all separate projectors, Eq. (5):

$$\hat{p} \equiv \prod_{i=1}^{N} \hat{p}_{s_i} \quad . \tag{5}$$

Because $\hat{p}_{s_i}$ is a linear combination of scalar products $\hat{\kappa}_{i,m} \cdot \hat{\kappa}_{i,n}$, $\hat{p}_{s_i}$ and $\hat{p}$ commute with the total spin, $[\hat{p}_{s_i}, \hat{\mathbf{S}}] = [\hat{p}, \hat{\mathbf{S}}] = 0$; thus, $[\hat{p}, \hat{P}_S] = 0$. Now consider a spin-1/2 configuration $|\phi\rangle$ that was selected for $\hat{P}_S$ using Löwdin's theorem. As mentioned, since $\hat{p}|\phi\rangle$ is symmetric under all permutations of the spin-1/2 centers within their respective sets of size $2s_i$, we can locally shift the $\downarrow$ in $|\phi\rangle$ all the way to the left. Consider, for example, four $s=1$ centers, each of which is split into two spin-1/2 components. A configuration selected for $S = M = 0$ according to Löwdin's theorem would be $|\phi_n\rangle = |\uparrow\downarrow\rangle|\uparrow\uparrow\rangle|\downarrow\downarrow\rangle|\uparrow\downarrow\rangle$. With respect to $\hat{p}$, $|\phi_n\rangle$ is equivalent to $|\phi'_n\rangle = |\downarrow\uparrow\rangle|\uparrow\uparrow\rangle|\downarrow\downarrow\rangle|\downarrow\uparrow\rangle$, with all $\downarrow$ moved to the left in their respective sets, see Eq. (6).

$$\hat{p}|\phi_n\rangle = \hat{p}|\phi'_n\rangle = |s_1 = s_2 = s_3 = s_4 = 1; m_1 = 0, m_2 = 1, m_3 = -1, m_4 = 0\rangle \tag{6}$$

However, $|\phi'_n\rangle$ does not fulfill the cumulative-sum criterion, so $\hat{P}_S|\phi'_n\rangle$ can be expressed as a linear combination of projected states that do satisfy $\sum_i^n m_i \geq 0$ for all $n$:

$$\hat{P}_S|\phi'_n\rangle = \sum_m c_m \hat{P}_S|\phi_m\rangle \quad . \tag{7}$$

We state without proof that $\hat{P}_S|\phi'_n\rangle \neq \hat{P}_S|\phi_n\rangle$. Thus, when rewriting Eq. (7) as Eq. (8), both sides are non-zero:

$$\hat{P}_S\left(|\phi'_n\rangle - c_n|\phi_n\rangle\right) = \sum_{m\neq n} c_m \hat{P}_S|\phi_m\rangle . \tag{8}$$

Now apply $\hat{p}$, using $[\hat{p},\hat{P}_S]=0$ and $\hat{p}|\phi_n\rangle = \hat{p}|\phi'_n\rangle$:

$$\begin{aligned}\hat{p}\hat{P}_S\left(|\phi'_n\rangle - c_n|\phi_n\rangle\right) &= \hat{P}_S\left(\hat{p}|\phi'_n\rangle - c_n\hat{p}|\phi_n\rangle\right) = \\ (1-c_n)\hat{P}_S\hat{p}|\phi'_n\rangle &= \sum_{m\neq n} c_m \hat{P}_S\hat{p}|\phi_m\rangle\end{aligned} \tag{9}$$

Eq. (9) shows that $\hat{P}_S\hat{p}|\phi'_n\rangle$ is a linear combination (with coefficients $\frac{c_m}{1-c_n}$, $m \neq n$) of $\hat{P}_S\hat{p}|\phi_m\rangle$, where the unprimed spin-1/2 configurations $|\phi_m\rangle$ obey Löwdin's theorem, and $|\phi'_n\rangle$ does not.

These considerations suggest a simple procedure to select the states $|m_1,...,m_N\rangle$ of the actual system, such that they span a complete and linearly independent set upon spin projection $\hat{P}_S$ ($S = M$): replace each $m_i$ with a configuration of $2s_i$ spin-1/2 centers, arranging a number of $s_i - m_i$ sites with ↓ on the left, and the remaining $s_i + m_i$ with ↑ on the right. The thus obtained configurations of all $N$ centers are concatenated into a single configuration of length $N\sum_i 2s_i$, and the cumulative-sum criterion, $\sum_i^n m_i \geq 0$, $n = 1,...,N\sum_i 2s_i$, as per Löwdin's theorem, is applied. A few examples are provided in the Results section. For systems with mixed local spin (i.e., not all $s_i$ are the same), contrary to the case of uniform $s$, the selected sets of configurations for different site numberings are in general not related by site permutations. However, each projected set completely spans the $S$ space, irrespective of the numbering. Although we have not presented a strict proof of the described procedure, numerous checks have confirmed its correctness.

One can straightforwardly check that a correct number $\dim(S)$ of basis states has been found by computing the difference between the dimensions of the $M = S$ and $M = S+1$ spaces, which are obtained by counting states with a respective $M = \sum_i m_i$. (A general formula for the dimensions of the $M$-spaces was derived in Ref. [20], and for numerous systems with uniform $s$, the dimensions $\dim(S)$ are collected in Table 1 of Ref. [12].)

Despite being linearly independent, the basis selected according to the (extended) Löwdin theorem is not necessarily optimally conditioned from the perspective of numerical stability. Particularly for larger systems, small eigenvalues of the overlap matrix can compromise the accuracy. Thus, as an alternative to selecting uncoupled basis functions following Löwdin's theorem, we use the pivoted Cholesky decomposition [21] (PCD), which is occasionally employed in quantum chemistry (see, e.g., Ref. [22], and references cited therein) to address the problem of over-complete basis sets by pruning them to yield optimal low-rank approximations, which enhances the numerical stability and efficiency of electronic-structure calculations. A PCD of the full overlap matrix $\mathbf{P}_S$ (see below) between spin-projected uncoupled states in a constant-$M$ space provides an optimal basis in terms of numerical stability but is generally not practical. Therefore we suggest the following iterative procedure: i) initial selection: spin-project a randomly selected set of $\dim(S)$ configurations and form the overlap matrix, ii) rank determination: calculate the rank $r$ of the overlap matrix within a tolerance well above the numerical accuracy threshold, iii) optimal subset: select $r$ uncoupled states through a PCD of the overlap matrix, iv) supplementary selection: if $r < \dim(S)$, add additional configurations to the selected $r$ states so that the total slightly exceeds $\dim(S)$, but avoid significantly exceeding $\dim(S)$, v) iteration: repeat the rank determination and state selection until $r$ has reached $\dim(S)$, vi) adjust numerical tolerance: if necessary, lower the tolerance for rank determination if $r$ remains too low even after considering significantly more than $\dim(S)$ states. The details of this procedure, or any similar approach, may be subject to optimization.

The selected uncoupled states form the columns of a matrix $\mathbf{R}$, i.e., each column of $\mathbf{R}$ contains a single entry 1. The construction of $\mathbf{H}$ in the uncoupled constant-$M$ basis is a standard task. For completeness, we shall briefly sketch it for $s = \frac{1}{2}$ (these considerations can be easily extended to $s > \frac{1}{2}$, see Ref. [23]). Each uncoupled state, e.g., $\left|\uparrow\downarrow...\uparrow\right\rangle$, is represented by a bit string of length $N$, where 0 at the $i$-th position denotes $\downarrow$ ($m_i = -\frac{1}{2}$) and 1 denotes $\uparrow$ ($m_i = +\frac{1}{2}$). For $J_{ij} = 1$ between nearest neighbors $\langle i,j \rangle$, the Hamiltonian is formulated in terms of raising and lowering operators in Eq. (10),

$$\hat{H} = \sum_{\langle i,j \rangle} \left[ \hat{s}_{i,z}\hat{s}_{j,z} + \tfrac{1}{2}\left(\hat{s}_{i,+}\hat{s}_{j,-} + \hat{s}_{i,-}\hat{s}_{j,+}\right) \right] \ , \tag{10}$$

where $\hat{s}_{i,z}\hat{s}_{j,z}$ contributes to the diagonal, and $\hat{s}_{i,-}\hat{s}_{j,+} = \left(\hat{s}_{i,+}\hat{s}_{j,-}\right)^{\dagger}$ flips the respective bits' values, thus accounting for off-diagonal elements. The number of non-zero entries in each row or column of $\mathbf{H}$ approximately corresponds to the number of interacting spin pairs, and $\mathbf{H}$ can be efficiently stored in sparse-matrix format [23]. As outlined below, we need to form the matrix product $\mathbf{R}^T\mathbf{H}$ (the superscript $T$ denotes transposition), and therefore limit ourselves to the rows corresponding to the selected configurations by directly constructing the rectangular matrix $\mathbf{R}^T \mathbf{H}$ instead of $\mathbf{H}$.

We now turn to the practical aspects of spin-adapting the basis. It is simplest to work directly with the Löwdin projector, Eq. (1). Since the spin square $\hat{\mathbf{S}}^2$ is the sum of the scalar products of all pairs, up to a constant (first sum in Eq. (11)),

$$\hat{\mathbf{S}}^2 = \sum_i s_i(s_i+1) + 2\sum_{i<j} \hat{\mathbf{s}}_i \cdot \hat{\mathbf{s}}_j \ , \tag{11}$$

its representation $\mathbf{S}^2$ can be calculated just as easily as $\mathbf{H}$, or the relation $\mathbf{S}^2 = \mathbf{S}_z^2 + \mathbf{S}_z + 2\mathbf{S}_-\mathbf{S}_+$ may be used [13]. Spin adaptation is formally accomplished by multiplying $\mathbf{R}$ with the projector, $\mathbf{P}_S \mathbf{R}$, although in practice one would successively apply the factors in the product of Eq. (1) to $\mathbf{R}$ instead of explicitly computing the dense matrix $\mathbf{P}_S$. Since the projector is Hermitian ($\mathbf{P}_S^\dagger = \mathbf{P}_S$) and idempotent ($\mathbf{P}_S \mathbf{P}_S = \mathbf{P}_S$) and commutes with the Hamiltonian, $[\mathbf{P}_S, \mathbf{H}] = 0$, it only appears once in the matrix products for the Hamiltonian and the overlap matrix (not to be confused with the spin vector), Eqs. (12) and (13), respectively.

$$\mathbf{H}_S = \mathbf{R}^\dagger \mathbf{P}_S^\dagger \mathbf{H} \mathbf{P}_S \mathbf{R} = \mathbf{R}^\dagger \mathbf{H} \mathbf{P}_S \mathbf{R} \tag{12}$$

$$\mathbf{S}_S = \mathbf{R}^\dagger \mathbf{P}_S \mathbf{R} \tag{13}$$

$\mathbf{H}_S$ and $\mathbf{S}_S$ define a generalized eigenvalue problem (GEP) whose solution yields the complete spectrum in the respective $S$ sector. This requires storing $\mathbf{S}_S$ in addition to $\mathbf{H}_S$. We solve the GEP with `Matlab`. Note that numerical rounding errors generally cause minor asymmetries in $\mathbf{H}_S$ and $\mathbf{S}_S$, which may lead to small imaginary components in the eigenvalues. We therefore perform a symmetrization, $\mathbf{H}_S = \frac{1}{2}(\mathbf{H}_S + \mathbf{H}_S^T)$ and $\mathbf{S}_S = \frac{1}{2}(\mathbf{S}_S + \mathbf{S}_S^T)$, which also significantly improves the computational efficiency of solving the GEP.

As indicated, one can alternatively write $\hat{P}_S$ in terms of an integration with respect to Euler angles $\alpha$, $\beta$ and $\gamma$,

$$\hat{P}_S = \frac{2S+1}{8\pi^2} \int_0^{2\pi} d\alpha \int_0^\pi d\beta \sin\beta \int_0^{2\pi} d\gamma \left[ D_{MM}^S(\alpha, \beta, \gamma) \right]^* e^{-i\alpha \hat{S}_z} e^{-i\beta \hat{S}_y} e^{-i\gamma \hat{S}_z} \ , \tag{14}$$

where $D_{MM}^S(\alpha, \beta, \gamma)$ is a diagonal element of the Wigner rotation matrix, and the asterisk * denotes complex conjugation. In a space with definite $M$, the group-theoretical projector,

Eq. (14), is fundamentally identical with Löwdin's operator, Eq. (1), but the former is preferred in methods like Projected Hartree–Fock [24,25] (PHF), where it simplifies the optimization of the reference state using self-consistent field or gradient-based procedures. PHF has also been applied to Heisenberg spin clusters [26–28]. In this approach, a symmetry-broken mean-field reference – which can be a product state of either individual spins [26] (such as a non-collinear spin configuration) or spin centers grouped into subclusters [27,28] – is optimized for spin- or PG-projection.

Here we would like to make use of the fact that the group-theoretical projector can be straightforwardly evaluated in closed form for $s = \frac{1}{2}$, as explained by Pauncz (Chapter 4.9 in Ref. [17]), and provide a compact derivation for completeness. Since we work with $\hat{S}_z$ eigenstates, the integrations over the Euler angles $\alpha$ and $\gamma$, both associated with rotations about the $z$-axis, can be directly evaluated [17,29], and yield factors that are irrelevant for our purpose. The non-trivial part of the projector thus reduces to an integration over $\beta$. Specifically, the wave function of $\hat{P}_S |m_i,...,m_N\rangle$ in the uncoupled basis ($S = M$) is formulated as an integral over products of elements of Wigner's small $d$-matrices in Eq. (15).

$$\langle m'_i,...,m'_N | \hat{P}_S | m_i,...,m_N \rangle \propto \langle m'_i,...,m'_N | \int d\beta\, d^S_{SS}(\beta) \sin\beta\, e^{-i\beta\hat{S}_y} | m_i,...,m_N \rangle = \int d\beta\, d^S_{SS}(\beta) \sin\beta \prod_{i=1}^{N} d^{s_i}_{m'_i m_i}(\beta) \quad (15)$$

The $d$-matrix for $s = \frac{1}{2}$ is given in Eq. (16).

$$\mathbf{d}^{1/2} = \begin{pmatrix} d^{1/2}_{1/2,1/2} & d^{1/2}_{1/2,-1/2} \\ d^{1/2}_{-1/2,1/2} & d^{1/2}_{-1/2,-1/2} \end{pmatrix} = \begin{pmatrix} \cos\frac{\beta}{2} & -\sin\frac{\beta}{2} \\ \sin\frac{\beta}{2} & \cos\frac{\beta}{2} \end{pmatrix} \quad (16)$$

Noting that $d^S_{SS}(\beta) \propto \cos^{2S}(\frac{\beta}{2})$, Eq. 15 represents a standard integral, Eq. (17), where $B$ is the Euler Beta-function, which can be expressed in terms of the Gamma-function $\Gamma$.

$$\int_0^{\pi} d\beta \sin^x\left(\frac{\beta}{2}\right) \cos^y\left(\frac{\beta}{2}\right) \sin\beta = 2B\left(\frac{x+2}{2}, \frac{y+2}{2}\right) = 2\frac{\Gamma\left(\frac{x+2}{2}\right)\Gamma\left(\frac{y+2}{2}\right)}{\Gamma\left(\frac{x+y+4}{2}\right)} \quad (17)$$

This leads to the result of Eq. (18),

$$\langle m'_i,...,m'_N | \hat{P}_S | m_i,...,m_N \rangle \propto C(N,S,k) \quad , \quad (18)$$

where $C(S,k)$ is a Sanibel-coefficient for the special case of $S = M$ [17],

$$C(N,S,k) = (-1)^k \frac{2S+1}{N_\uparrow +1} \binom{N_\uparrow}{k}^{-1}. \tag{19}$$

$C(N,S,k)$ is essentially a sign factor divided by a binomial coefficient; $k$ counts the number of sites with $m_i' < m_i$, and $N_\uparrow$ is the number of $\uparrow$ sites, i.e., $N_\uparrow + N_\downarrow = N$, $M = \frac{1}{2}(N_\uparrow - N_\downarrow)$. As explained, the proportionality factor in Eq. (18) is not relevant here, and we omit it to avoid clutter.

For $s_i > \frac{1}{2}$, the closed-form solution of the integral in Eq. (15) in general represents a linear combination of standard integrals, Eq. (17), because the $d^{s_i}_{m_i'm_i}(\beta)$ with $|m_i| < s_i$ and $|m_i'| < s_i$ are linear combinations of $\sin^a(\frac{\beta}{2})\cos^b(\frac{\beta}{2})$ with different sets of (real integer) exponents $(a, b)$, see Eqs. (20) and (21) for $s = 1$ and $s = \frac{3}{2}$, respectively, where $s \equiv \sin\frac{\beta}{2}$ and $c \equiv \cos\frac{\beta}{2}$.

$$\mathbf{d}^1 = \begin{pmatrix} d^1_{1,1} & d^1_{1,0} & d^1_{1,-1} \\ d^1_{0,1} & d^1_{0,0} & d^1_{0,-1} \\ d^1_{-1,1} & d^1_{-1,0} & d^1_{-1,-1} \end{pmatrix} = \begin{pmatrix} c^2 & -\sqrt{2}sc & s^2 \\ \sqrt{2}sc & c^2 - s^2 & -\sqrt{2}sc \\ s^2 & \sqrt{2}sc & c^2 \end{pmatrix} \tag{20}$$

$$\mathbf{d}^{3/2}(\beta) = \begin{pmatrix} d^{3/2}_{3/2,3/2} & d^{3/2}_{3/2,1/2} & d^{3/2}_{3/2,-1/2} & d^{3/2}_{3/2,-3/2} \\ d^{3/2}_{1/2,3/2} & d^{3/2}_{1/2,1/2} & d^{3/2}_{1/2,-1/2} & d^{3/2}_{1/2,-3/2} \\ d^{3/2}_{-1/2,3/2} & d^{3/2}_{-1/2,1/2} & d^{3/2}_{-1/2,-1/2} & d^{3/2}_{-1/2,-3/2} \\ d^{3/2}_{-3/2,3/2} & d^{3/2}_{-3/2,1/2} & d^{3/2}_{-3/2,-1/2} & d^{3/2}_{-3/2,-3/2} \end{pmatrix} =$$
$$\begin{pmatrix} c^3 & -\sqrt{3}c^2 s & \sqrt{3}cs^2 & -s^3 \\ \sqrt{3}c^2 s & c^3 - 2cs^2 & -2c^2 s + s^3 & \sqrt{3}cs^2 \\ \sqrt{3}cs^2 & 2c^2 s - s^3 & c^3 - 2cs^2 & -\sqrt{3}c^2 s \\ s^3 & \sqrt{3}cs^2 & \sqrt{3}c^2 s & c^3 \end{pmatrix} \tag{21}$$

By taking these linear combinations into account, we can still evaluate the integral in closed form, which we have implemented up to $s = \frac{3}{2}$. Note that, in general, the combinatorial growth of the length of these linear combinations with the number of spin centers may be mitigated by selecting the uncoupled basis through PCD. For example, for $N = 8$, $s = 1$, we found that PCD based on the full overlap matrix $\mathbf{P}_S$ in the $S = 0$ sector yields only Ising-type configurations with local projections $m = \pm 1$. Since $m = 0$ is thereby excluded, the integral for the group-theoretical projection is just a single term as for $s = \frac{1}{2}$ systems. Thus, in a limited PCD procedure, as explained above, one may preferentially select configurations with only a few local $z$-projections that do not have their maximal magnitude to keep linear combinations in closed-form integral solutions short. However, optimizing such a procedure is beyond the scope of this work.

Alternatively, we can discretize the integration using a Gauss-Legendre grid [29,30]. This scales approximately linearly with the number of grid points (larger systems generally require larger grids), and for $s = \frac{1}{2}$ it is significantly less efficient than using the closed-form solution of Eq. (18). In the Results section, we illustrate some examples using integration on a grid. In our `MATLAB` code, we have embedded several tasks, including the integration of products of Wigner-$d$-matrix elements (Eq. (15)), with `mex-C` functions for efficiency. For $s > \frac{1}{2}$, projection using either Löwdin's operator, or discretized or exact SU(2) integration all require similar amounts of computation time. However, we cannot rule out that in an optimal implementation, one of these options might be significantly advantageous.

## 3. Results and Discussion

Here, we first present examples for selecting basis states according to the extended Löwdin theorem and then compare the various presented options for spin projection in terms of their numerical accuracy by calculating spectra of antiferromagnetic rings ($J = 1$ between nearest neighbors).

Figure 2 illustrates the cumulative sum of the local $z$-projections as a function of the number $n$ of auxiliary spin-1/2 sites for a system with eight $s = \frac{3}{2}$ centers. Each path in the diagram, moving from left to right, represents a configuration selected by the extension of Löwdin's theorem. When $\hat{P}_S$ is applied to these configurations, they form a complete, linearly independent basis for the given $S$. In each panel, an arbitrarily chosen configuration is highlighted for illustration, which is expressed as a product of states from three spin-1/2 centers, each representing a single $s = \frac{3}{2}$. This spin-1/2 configuration is translated back into a state of the actual system by summing the $z$-projections of the respective spin-1/2 components. As explained, within each center, if ↓ sites are present, they are locally positioned to the left; thus, a configuration is only compatible if $m_1 = s_1$.

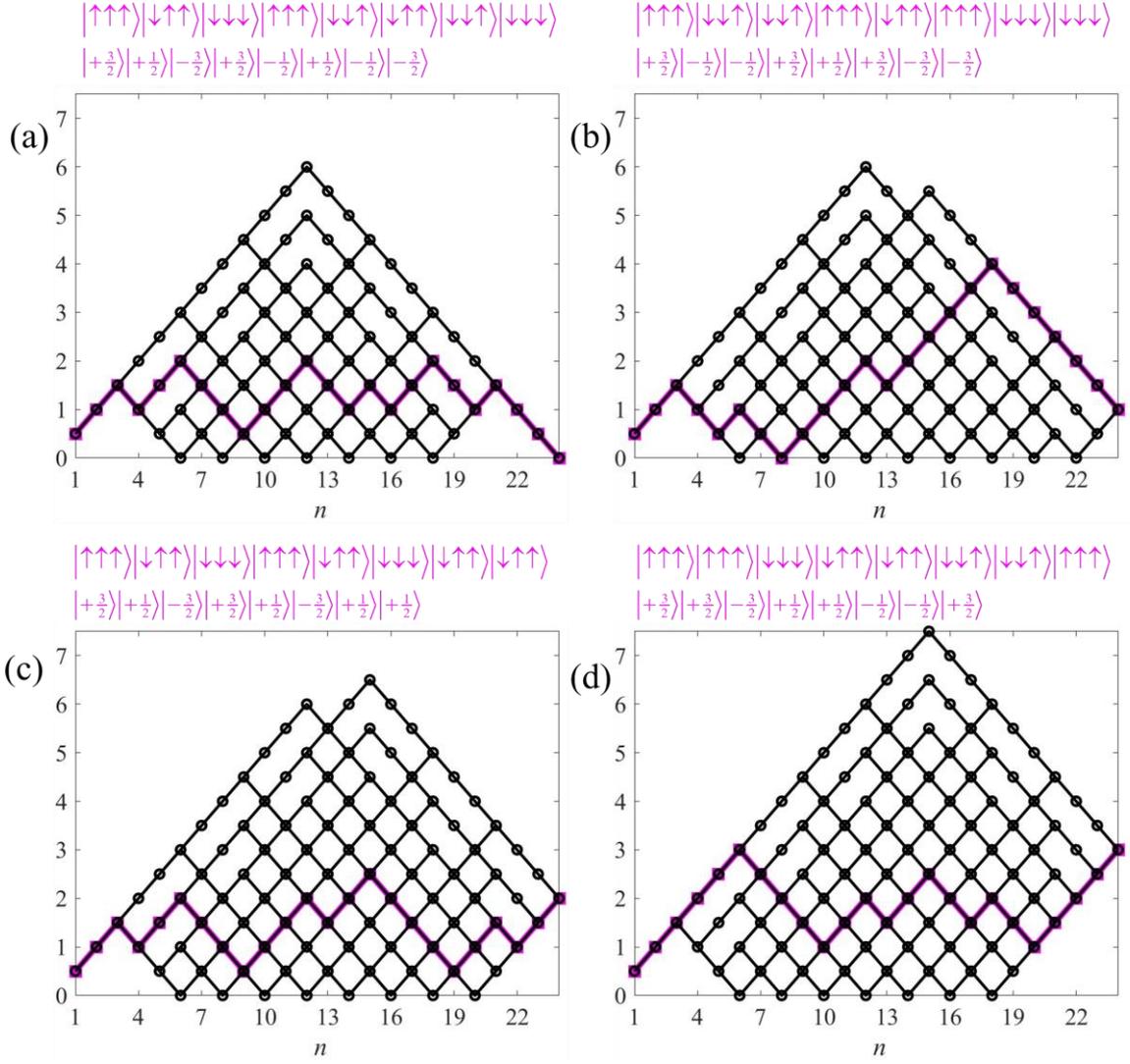

Figure 2: Cumulative sum of the local $z$-projections, $\sum_{i=1}^{n} m_i$, as a function of the number $n$ of auxiliary spin-1/2 sites for a system with eight $s = \frac{3}{2}$ centers. Each path in the diagram, moving from left to right, represents an uncoupled configuration of the system selected by the extension of Löwdin's theorem, for projection onto sectors $S = 0$ (a), $S = 1$ (b), $S = 2$ (c) and $S = 3$ (d). For further details, see main text.

Figure 3 shows the energy levels of the antiferromagnetic $N = 12$ spin-1/2 ring as a function of $S$. Our reference for assessing the numerical accuracy of the different projection schemes are the exact energy levels $E_{ex}$ from full ED of **H**. By comparing the spectra across all different $M$-spaces, each level is assigned a spin $S$. Since **H** is constructed numerically exactly, i.e., without (or with negligible) rounding errors, these eigenvalues are accurate within double precision in `Matlab`, corresponding to approximately 15 or 16 decimal places. In Figure 4, we plot the logarithmic difference between the numerically exact levels and the levels obtained from the GEP that was set up through either the Löwdin-projector, Gauss-Legendre integration with $g = 12$ grid points, or Sanibel coefficients. The grid integration is employed here mainly for comparison with the preferable use of Sanibel coefficients, which, as explained in the Theory section, result from a closed-form solution of the SU(2) integral.

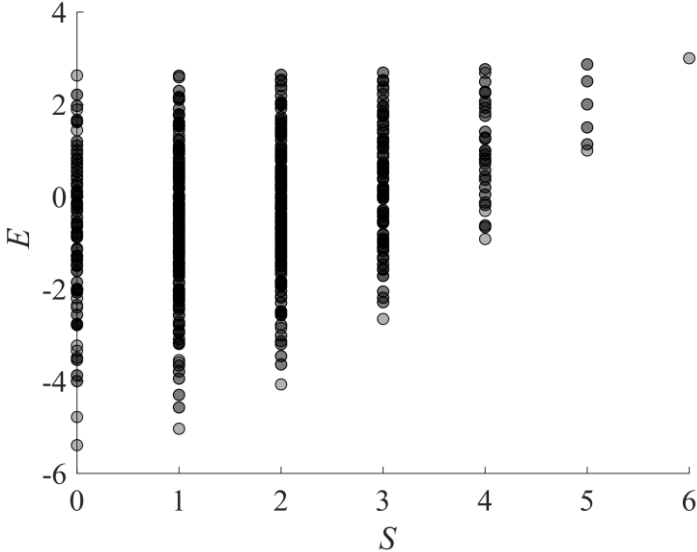

Figure 3: Full spectrum of an antiferromagnetic $N = 12$ spin-1/2 ring (coupling constant $J = 1$), as a function of the total spin $S$.

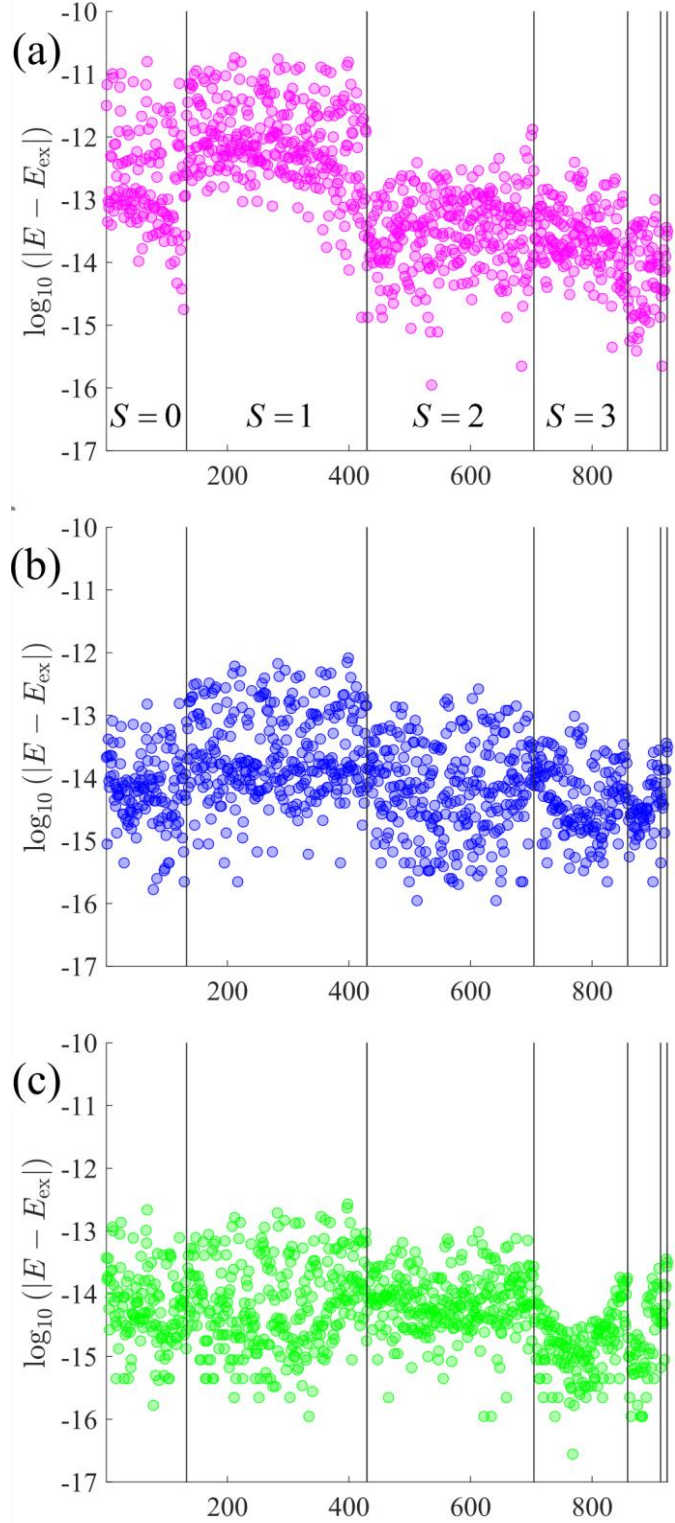

Figure 4: Numerical accuracy of three different projection schemes for an antiferromagnetic $N = 12$ spin-1/2 ring. The logarithmic difference between the exact energies and the generalized eigenvalues is plotted. The GEP was constructed based on the Löwin projector (a), a Gauss-Legendre integration with $g = 12$ points (b), or Sanibel-coefficients (c). Data points corresponding to different $S$ are separated by vertical lines, and within each $S$ sector, the energies are ordered in ascending fashion from left to right.

All three methods are accurate to within $< 10^{-10}$, with the maximum error being greatest for Löwdin's projector. The $g = 12$ grid size is sufficient. However, for $g = 4$ a peculiar pattern emerges for $S \geq 2$: some states are described accurately, while others exhibit significant errors, resulting in a large gap between the two groups, see Figure 5. Interestingly, all states with $S = 3$ or $S = 4$ belong to latter group, whereas for $S = 2$ or $S = 5$, some levels are accurate and others are not.

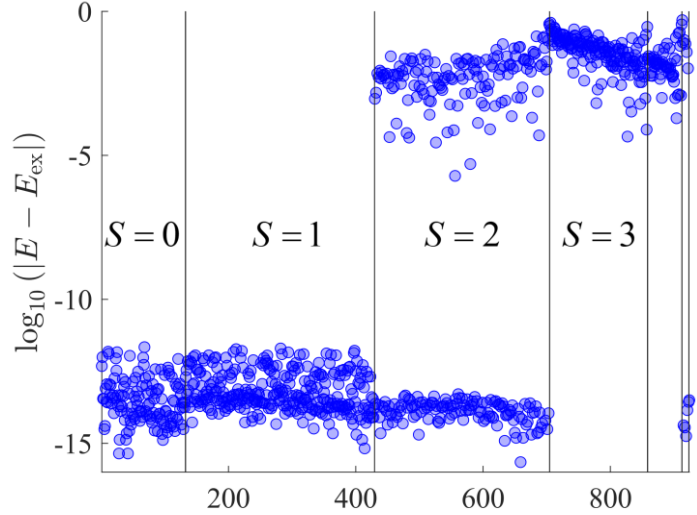

Figure 5: Numerical accuracy of $g = 4$ Gauss-Legendre integration for an antiferromagnetic $N = 12$ spin-1/2 ring. For further details, see caption to Figure 4 and main text.

Figure 4 illustrates the corresponding numerical results for an $N = 16$ ring. Again, Löwdin's projector is less accurate compared to using Sanibel coefficients. Notably, the maximum error for eigenvalues obtained based on Löwdin's projector is by almost four orders of magnitude larger than for the $N = 12$ ring. Projecting out seven contaminating spin contributions ($l \neq S$ in Eq. (1)) from each configuration, as opposed to five for $N = 12$, requires more matrix-vector products and thus leads to a greater accumulation of rounding errors. Even the projection using Sanibel coefficients now shows larger errors, although these remain $< 10^{-10}$.

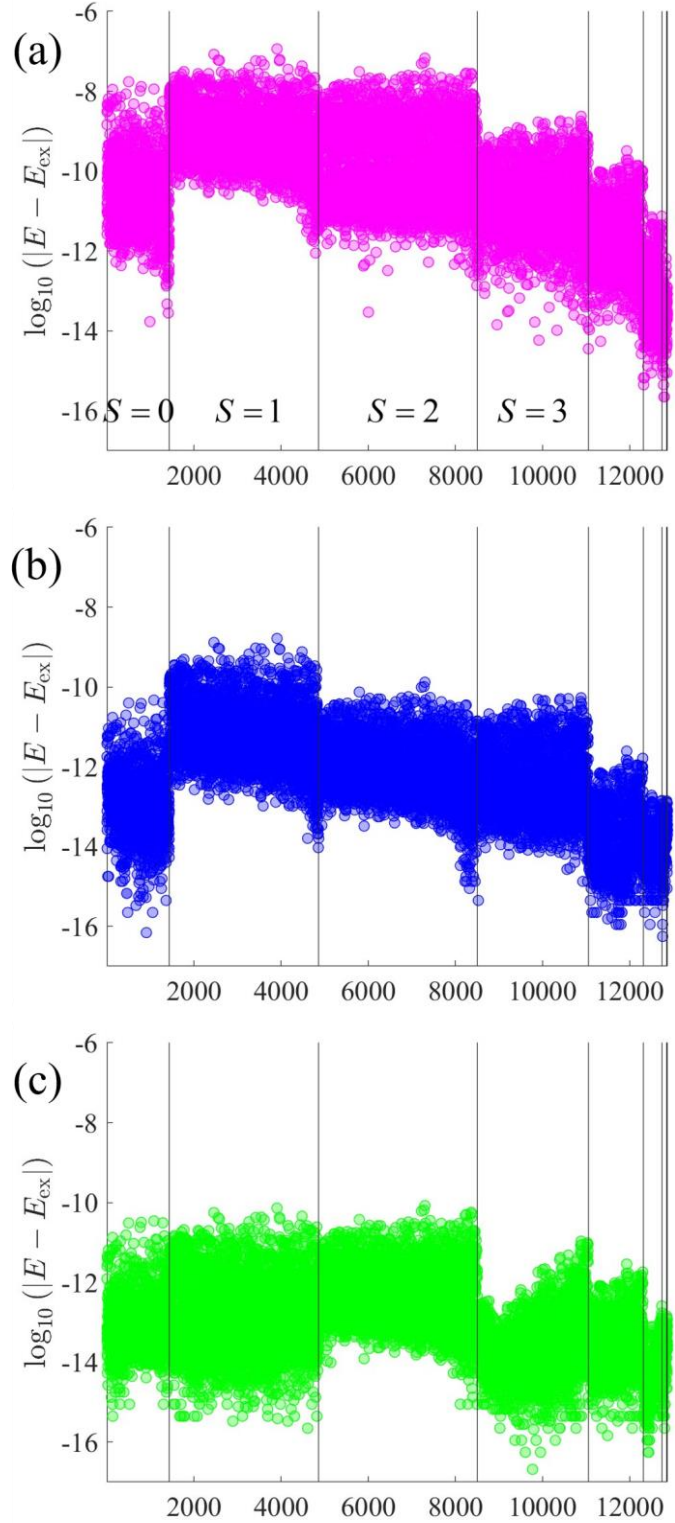

Figure 6: Numerical accuracy of different projection schemes (a: Löwdin, b: integration with $g = 12$ grid points, c: Sanibel coefficients) for an antiferromagnetic $N = 16$ spin-1/2 ring. For further details, see caption to Figure 4.

Finally, as an example of an $s > \frac{1}{2}$ system, we choose a ring with $s = \frac{3}{2}$ and $N = 8$. The full spectrum is shown in Figure 7.

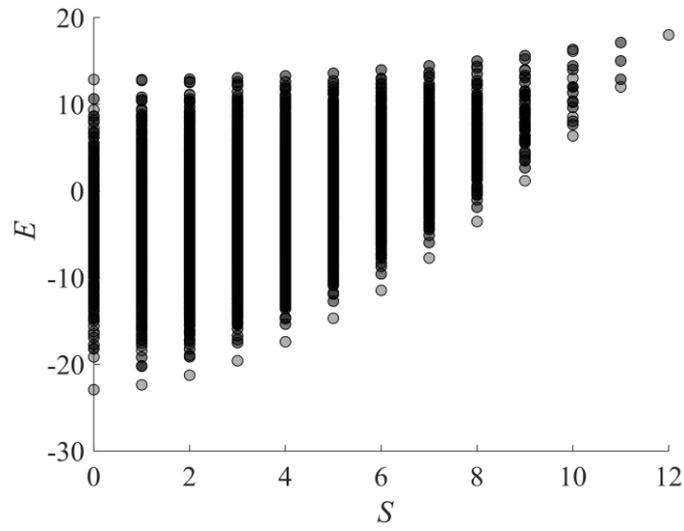

Figure 7: Full spectrum of an antiferromagnetic $N = 8$ $s = \frac{3}{2}$ ring, $J = 1$, as a function of $S$.

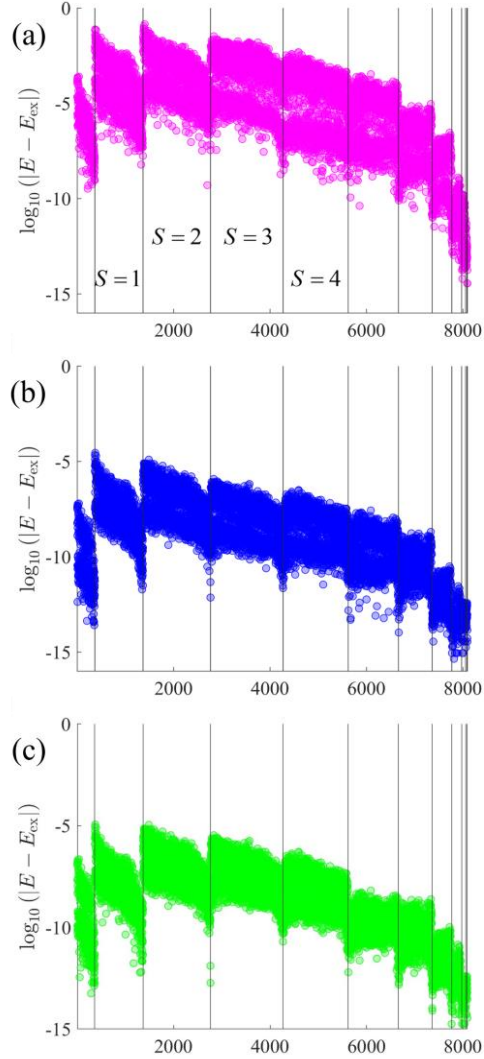

Figure 8: Numerical accuracy of different projection schemes (a: Löwdin, b: integration with $g = 12$ grid points, c: exact integration using a closed-form solution to Eq. (15)) for an antiferromagnetic $N = 8$ $s = \frac{3}{2}$ ring. For further details, see caption to Figure 4.

Figure 8 shows that grid-integration yields significantly more accurate spectra than Löwdin's projector. Using the latter, the maximal error is $|E - E_{ex}| = 0.143$ (in units of the coupling constant $J = 1$). Such large deviations could become noticeable when fitting inelastic neutron scattering (INS) spectra or thermodynamic data such as magnetic susceptibilities. Bernu et al [13]. have already pointed out that rounding errors can quickly accumulate with Löwdin's projector, necessitating additional mitigation measures. In this regard, an advantage of the group-theoretical formulation becomes evident. However, errors remain comparatively large even when the integrals of Eq. (15) are evaluated exactly (based on a closed-form solution of Eq. (15)), see Figure 8c. The limited accuracy indeed derives from near linear dependencies in the spin-projected basis.

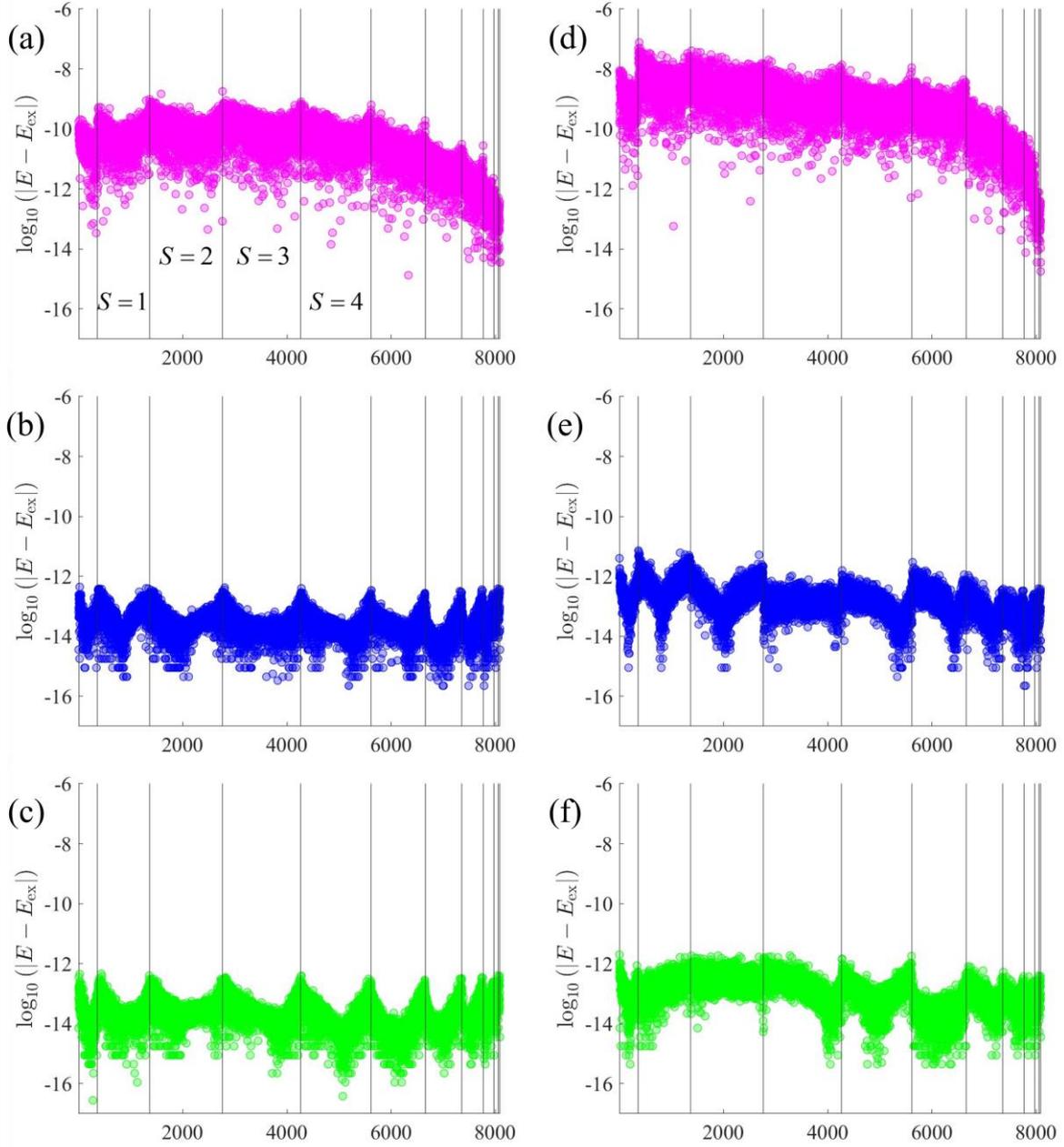

Figure 9: Numerical accuracy of different projection schemes (a, d: Löwdin, b, e: integration with $g = 12$ grid points, c, f: exact integration using a closed-form solution to Eq. (15)) for an antiferromagnetic $N = 8$ $s = \frac{3}{2}$ ring based on uncoupled states selected by either a PCD of the full overlap matrix $\mathbf{P}_S$ (a, b, c) or using an iterative scheme (d, e, f). For further details, see caption to Figure 4.

This is demonstrated by using PCD to select the uncoupled states. The left panel of Figure 9 depicts results based on the impractical PCD of the full overlap matrix, while the right panel was obtained from basis selection through the described iterative PCD procedure that considers only a randomly selected fraction of states. It is evident that Löwdin's projector still affords larger errors than the group-theoretical formulation, but all results are significantly more accurate than when selecting states according to the extended Löwdin theorem.

## 4. Summary and Outlook

The conventional approach for spin adaptation to facilitate the diagonalization of isotropic spin Hamiltonians requires some familiarity with irreducible tensor operators (ITOs) and spin-coupling techniques, and its implementation is therefore comparatively demanding. In contrast, the spin projection of uncoupled configurations requires little more than the ability to compute the Hamiltonian in an uncoupled basis, which is a standard task. Irrespective of the formulation of the projector, one can select uncoupled configurations so that a complete, linearly independent basis is spanned by projection. To this end, we have found a simple and intuitive extension of Löwdin's theorem from $s = \frac{1}{2}$ to any local spin $s$, based on expanding configurations in terms of auxiliary spin-1/2 sites. However, the conditioning of the overlap matrix of spin-projected states formed from these configurations could still become problematic in larger systems, and we thus suggest to iteratively apply a pivoted Cholesky decomposition for better numerical stability in such cases.

Löwdin's projector can be constructed straightforwardly. The accumulation of rounding errors in the formation of successive matrix-vector products for each spin component to be projected out ($l \neq S$ in Eq. (1)) is typically not practically significant, but would need to be mitigated beyond a certain system size. On the other hand, the fact that spin-projection for $s = \frac{1}{2}$ can also be achieved using simple Sanibel coefficients, which result from the closed-form evaluation of the group-theoretical projection operator, seems to have gone unnoticed for the diagonalization of Heisenberg clusters. This strategy is not only computationally simpler but also advantageous in terms of numerical accuracy. We have also solved the SU(2) integral for $s > \frac{1}{2}$ in closed form, which generally involves a linear combination of terms corresponding to standard integrals. Alternatively, the integral may be approximated on a grid in terms of a weighted sum of products of Wigner-$d$-matrix elements.

While the conventional spin-coupling ITO-method directly yields an orthogonal basis, the projection of uncoupled states results in a linearly independent but nonorthogonal basis. Consequently, in addition to the Hamiltonian, one needs to store an overlap matrix, and solving a generalized eigenvalue problem (GEP) takes slightly longer compared to an ordinary (orthogonal) EP. However, we believe that the advantages of the simpler practical realization outweigh this minor drawback. Note that we have not conducted a comparison of computational efficiency between ITO and projection methods, as an objective assessment would have to ensure that both methods are implemented with similar levels of optimization in a specific programming language, etc.

Spin clusters often exhibit high spatial symmetry, and the Hamiltonian is maximally factored only when spin as well as point-group (PG) symmetry are utilized. The simultaneous spin- and PG-adaptation is considered a challenging task because the spin-coupling method faces the problem that PG-operations correspond to complicated recoupling transformations [8,31–33]. On the other hand, using spin- and PG-projectors on uncoupled states [13,14] is hardly any more complicated than the pure spin-adaptation presented here. We plan to employ combined spin- and PG-projection to some of the largest fully diagonalizable Heisenberg spin clusters in a future work.

**References**

1. Caneschi, A.; Gatteschi, D.; Sessoli, R.; Barra, A.L.; Brunel, L.C.; Guillot, M. Alternating Current Susceptibility, High Field Magnetization, and Millimeter Band EPR Evidence for a Ground $S$ = 10 State in $[Mn_{12}O_{12}(CH_3COO)_{16}(H_2O)_4]\cdot 2CH_3COOH\cdot 4H_2O$. *J. Am. Chem. Soc.* **1991**, *113*, 5873.

2. Sessoli, R.; Gatteschi, D.; Caneschi, A.; Novak, M.A. Magnetic Bistability in a Metal-Ion Cluster. *Nature* **1993**, *365*, 141.

3. Bencini, A.; Gatteschi, D. *Electron Paramagnetic Resonance of Exchange Coupled Systems*; Springer: Berlin/Heidelberg, 1990.

4. Schnack, J. Large Magnetic Molecules and What We Learn from Them. *Contemp. Phys.* **2019**, *60*, 127.

5. Gatteschi, D.; Sessoli, R. Quantum Tunneling of Magnetization and Related Phenomena in Molecular Materials. *Angew. Chemie Int. Ed.* **2003**, *42*, 268.

6. Waldmann, O.; Güdel, H.U. Many-Spin Effects in Inelastic Neutron Scattering and Electron Paramagnetic Resonance of Molecular Nanomagnets. *Phys. Rev. B* **2005**, *72*, 94422.

7. Ghassemi Tabrizi, S.; Arbuznikov, A. V; Kaupp, M. Construction of Giant-Spin Hamiltonians from Many-Spin Hamiltonians by Third-Order Perturbation Theory and Application to an $Fe_3Cr$ Single-Molecule Magnet. *Chem. Eur. J.* **2016**, *22*, 6853.

8. Waldmann, O. Symmetry and Energy Spectrum of High-Nuclearity Spin Clusters. *Phys. Rev. B* **2000**, *61*, 6138.

9. Tsukerblat, B.S. *Group Theory in Chemistry and Spectroscopy*, 2nd ed.; Dover Publications: New York, 2006.

10. Borrás-Almenar, J.J.; Clemente-Juan, J.M.; Coronado, E.; Tsukerblat, B.S. MAGPACK 1 A Package to Calculate the Energy Levels, Bulk Magnetic Properties, and Inelastic Neutron Scattering Spectra of High Nuclearity Spin Clusters. *J. Comput. Chem.* **2001**, *22*, 985.

11. Chilton, N.F.; Anderson, R.P.; Turner, L.D.; Soncini, A.; Murray, K.S. PHI: A Powerful New Program for the Analysis of Anisotropic Monomeric and Exchange-Coupled Polynuclear *d*-and *f*-Block Complexes. *J. Comput. Chem.* **2013**, *34*, 1164.

12. Boča, R.; Rajnák, C.; Titiš, J. Spin Symmetry in Polynuclear Exchange-Coupled Clusters. *Magnetochemistry* **2023**, *9*, 226.

13. Bernu, B.; Lecheminant, P.; Lhuillier, C.; Pierre, L. Exact Spectra, Spin Susceptibilities, and Order Parameter of the Quantum Heisenberg Antiferromagnet on the Triangular Lattice. *Phys. Rev. B* **1994**, *50*, 10048.

14. Ghassemi Tabrizi, S.; Kühne, T.D. Analytical Solutions of Symmetric Isotropic Spin Clusters Using Spin and Point Group Projectors. *Magnetism* **2024**, *4*, 183.

15. Löwdin, P.-O. Angular Momentum Wavefunctions Constructed by Projector Operators. *Rev. Mod. Phys.* **1964**, *36*, 966.

16. Gershgorn, Z. Proof of the Linear Independence of Properly Selected Projected Spin Eigenfunctions. *Int. J. Quantum Chem.* **1968**, *2*, 341.

17. Pauncz, R. Spin Eigenfunctions: Construction and Use; Springer US: Boston, MA, 1979.

18. Pauncz, R. *The Construction of Spin Eigenfunctions: An Exercise Book*; Springer Science & Business Media, 2012.

19. Löwdin, P.-O. Quantum Theory of Many-Particle Systems. III. Extension of the Hartree–Fock Scheme to Include Degenerate Systems and Correlation Effects. *Phys. Rev.* **1955**, *97*, 1509.

20. Bärwinkel, K.; Schmidt, H.-J.; Schnack, J. Structure and Relevant Dimension of the Heisenberg Model and Applications to Spin Rings. *J. Magn. Magn. Mater.* **2000**, *212*, 240.

21. Harbrecht, H.; Peters, M.; Schneider, R. On the Low-Rank Approximation by the Pivoted Cholesky Decomposition. *Appl. Numer. Math.* **2012**, *62*, 428.

22. Lehtola, S. Curing Basis Set Overcompleteness with Pivoted Cholesky Decompositions. *J. Chem. Phys.* **2019**, *151*, 241102.

23. Raghu, C.; Rudra, I.; Sen, D.; Ramasesha, S. Properties of Low-Lying States in Some High-Nuclearity Mn, Fe, and V Clusters: Exact Studies of Heisenberg Models. *Phys. Rev. B* **2001**, *64*, 064419.

24. Schmid, K.W.; Dahm, T.; Margueron, J.; Müther, H. Symmetry-Projected Variational Approach to the One-Dimensional Hubbard Model. *Phys. Rev. B* **2005**, *72*, 85116.

25. Jiménez-Hoyos, C.A.; Henderson, T.M.; Tsuchimochi, T.; Scuseria, G.E. Projected Hartree–Fock Theory. *J. Chem. Phys.* **2012**, *136*, 164109.

26. Ghassemi Tabrizi, S.; Jiménez-Hoyos, C.A. Ground States of Heisenberg Spin Clusters from Projected Hartree–Fock Theory. *Phys. Rev. B* **2022**, *105*, 35147.

27. Ghassemi Tabrizi, S.; Jiménez-Hoyos, C.A. Ground States of Heisenberg Spin Clusters from a Cluster-Based Projected Hartree–Fock Approach. *Condens. Matter* **2023**, *8*, 18.

28. Papastathopoulos-Katsaros, A.; Henderson, T.M.; Scuseria, G.E. Symmetry-Projected Cluster Mean-Field Theory Applied to Spin Systems. *J. Chem. Phys.* **2023**, *159*, 084107.

29. Rivero, P.; Jiménez-Hoyos, C.A.; Scuseria, G.E. Entanglement and Polyradical Character of Polycyclic Aromatic Hydrocarbons Predicted by Projected Hartree–Fock Theory. *J. Phys. Chem. B* **2013**, *117*, 12750.

30. Ghassemi Tabrizi, S.; Arbuznikov, A. V; Jimenez-Hoyos, C.A.; Kaupp, M. Hyperfine-Coupling Tensors from Projected Hartree-Fock Theory. *J. Chem. Theory Comput.* **2020**, *16*, 6222.

31. Schnalle, R.; Schnack, J. Calculating the Energy Spectra of Magnetic Molecules: Application of Real-and Spin-Space Symmetries. *Int. Rev. Phys. Chem.* **2010**, *29*, 403.

32. Delfs, C.; Gatteschi, D.; Pardi, L.; Sessoli, R.; Wieghardt, K.; Hanke, D. Magnetic Properties of an Octanuclear Iron (III) Cation. *Inorg. Chem.* **1993**, *32*, 3099.

33. Schnalle, R.; Schnack, J. Numerically Exact and Approximate Determination of Energy Eigenvalues for Antiferromagnetic Molecules Using Irreducible Tensor Operators and General Point-Group Symmetries. *Phys. Rev. B* **2009**, *79*, 104419.